\documentclass[runningheads]{llncs}

\usepackage[T1]{fontenc}
\def\doi#1{\href{https://doi.org/\detokenize{#1}}{\url{https://doi.org/\detokenize{#1}}}}
\usepackage{graphicx}
\usepackage{makecell, booktabs}

\usepackage{listings}
\begin{document}

\title{CACTUSS: Common Anatomical CT-US Space for US examinations}

\author{Yordanka Velikova\inst{1} \and
 Walter Simson\inst{1} \and
 Mehrdad Salehi\inst{1} \and
 Mohammad Farid Azampour\inst{1,3} \and
 Philipp Paprottka\inst{2} \and
 Nassir Navab\inst{1,4}
}
%1{Velikova, Yordanka} 
%2{Simson, Walter} 
%3{Salehi, Mehrdad} 
%4{Azampour, Mohammad Farid} 
%5{Paprottka, Philipp} 
%6{Navab, Nassir} 

\authorrunning{Y. Velikova et al.}
 \institute{Computer Aided Medical Procedures, Technical University of Munich, Germany \\ 
 \and
 Interventional Radiology, Klinikum rechts der Isar, Munich, Germany \\ 
 \and
 Department of Electrical Engineering, Sharif University of Technology, Tehran, Iran \\ 
 \and
 Computer Aided Medical Procedures, John Hopkins University, Baltimore, USA \\
 }
\titlerunning\space{CACTUSS: Common Anatomical CT-US Space}
\maketitle              % typeset the header of the contribution
\begin{abstract}

Abdominal aortic aneurysm (AAA) is a vascular disease in which a section of the aorta enlarges, weakening its walls and potentially rupturing the vessel.
Abdominal ultrasound has been 
utilized for diagnostics, but due to its limited image quality and operator dependency, CT scans are usually required for monitoring and treatment planning.
Recently, abdominal CT datasets have been successfully utilized to train deep neural networks for automatic aorta segmentation. Knowledge gathered from this solved task could therefore be leveraged to improve US segmentation for AAA diagnosis and monitoring.
To this end, we propose CACTUSS: a common anatomical CT-US space, which acts as a virtual bridge between CT and US modalities to enable automatic AAA screening sonography. 
CACTUSS makes use of publicly available labelled data to learn to segment based on an intermediary representation that inherits properties from both US and CT. 
We train a segmentation network in this new representation and employ an additional image-to-image translation network which enables our model to perform on real B-mode images.
Quantitative comparisons against fully supervised methods demonstrate the capabilities of CACTUSS in terms of Dice Score and diagnostic metrics, showing that our method also meets the clinical requirements for AAA scanning and diagnosis.

\keywords{Ultrasound  \and Computer Aided Intervention \and Abdominal Aortic Aneurysm \and Domain Adaptation}

\end{abstract}
\section{Introduction}%1 page

Abdominal aortic aneurysm (AAA) is a life-threatening disease of the main blood vessel in the human body, the aorta, where an aneurysm, or expansion, occurs thereby weakening the aorta walls.
AAA can lead to a high risk of a rupturing of the aorta with an overall incidence rate of 1.9\% to 18.5\%, in males age 60+ years of age and an average subsequent mortality rate 60\%~\cite{ullery_epidemiology_2018}.

Abdominal ultrasound has been recommended as an initial examination modality for asymptomatic patients with a high risk of AAA.
There is evidence of a significant reduction of premature death from AAA in men aged 65 and above who undergo ultrasound screening~\cite{ullery_epidemiology_2018}.
Per definition, the aorta is considered aneurysmatic when the absolute anterior to posterior diameter is larger than 3~cm, independently of the relative body size of the patient.
\begin{figure}[t]
\includegraphics[width=\textwidth]{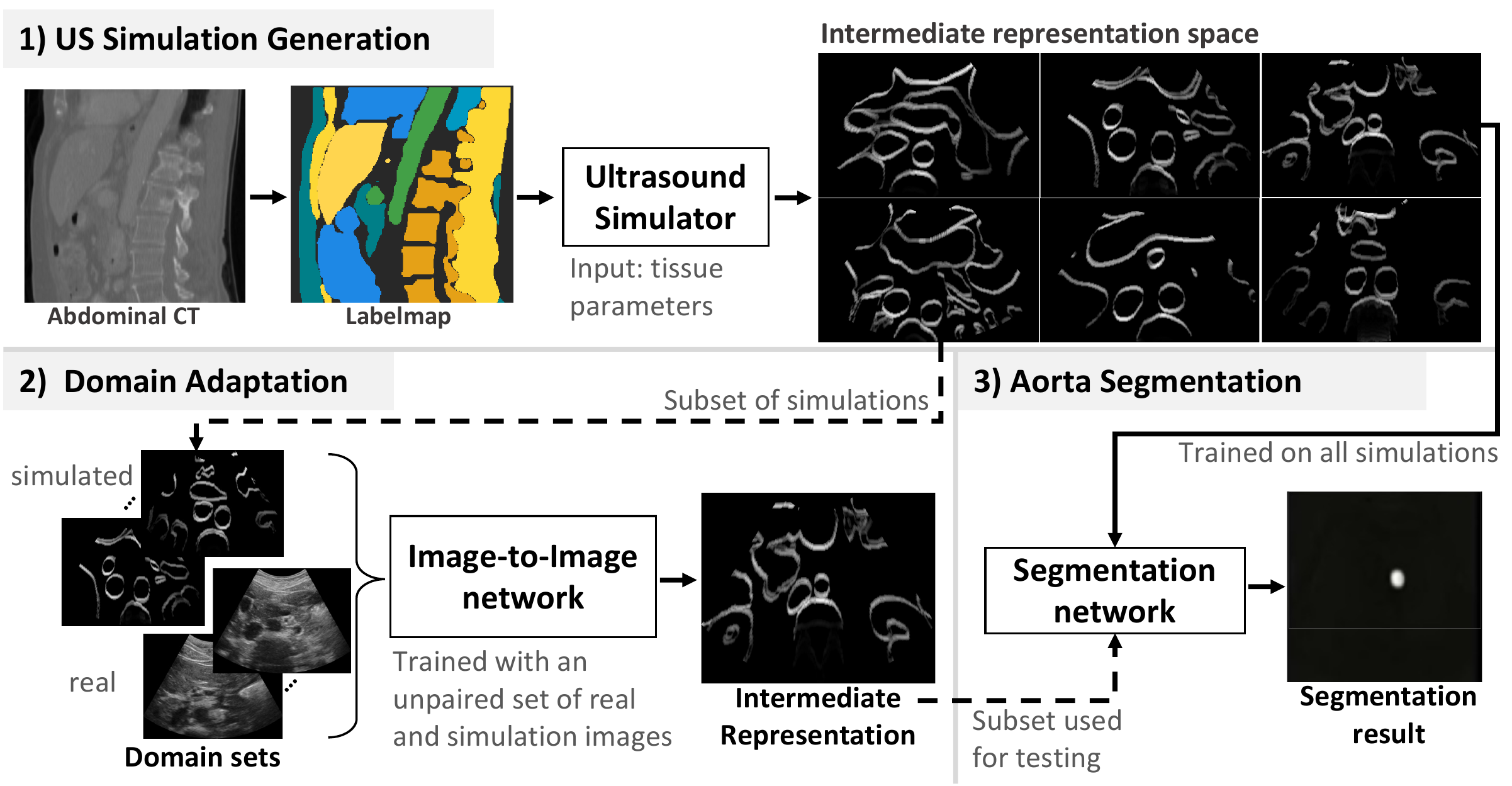}
\caption{Overview of the proposed framework. In phase one, an established ultrasound simulator is re-purposed and parameterized to define an intermediate representation, between the ultrasound and CT space. In phase two, an unsupervised network is trained separately in an isolated fashion to translate clinical ultrasound images to the intermediate representation defined in phase one. In phase three, a segmentation network is trained on the segmentation task using only samples generated with the ultrasound simulator. At inference time, real ultrasound images are passed to the image-to-image network, translated to the intermediate representation and segmented. This is the first time that the segmentation network has seen the intermediate representation from real ultrasound images.  } \label{fig:overview}
\end{figure}
However, because the interpretation of the US image is heavily based on the sonographer's experience, the resulting diagnosis is largely operator-dependent, as reported in~\cite{diagnosticUS}.

To overcome the challenge of reproducible ultrasound screening, robotic ultrasound (RUS) imaging has been proposed to offer reproducible  ultrasound scans independent of operator skill~\cite{merouche2015robotic,jiang2021autonomous,kojcev2017reproducibility}.
Specifically for screening of AAA, this has required an external camera and MRI atlas to locate and track the trajectory of the aorta, which reduces the usability and subsequent acceptance of the methods~\cite{virga2016automatic,langsch_robotic_2019}. Furthermore, ultrasound image quality has been criticized for not offering the resolution needed to make an accurate measurement~\cite{HARTSHORNE2011195}.

Computed tomography (CT) scans are  used in clinical practice to assess, manage, and monitor AAA after an initial discovery during screening~\cite{CHAIKOF20182}.
In recent years segmentation models based on deep neural networks have demonstrated great performance for automatized CT aorta segmentation, and numerous studies have been trained on large, expert annotated, and publicly available datasets~\cite{annurev_bioeng,brutti,LOPEZLINARES,CAO2019108713}.
This leads to the possible application of automatic screening and monitoring of AAA in CT imaging using deep learning~\cite{Yamashita2018ConvolutionalNN}.
However, acquiring a CT scan exposes the patient to ionizing radiation.

Ultrasound imaging can serve as a viable alternative to CT and help to reduce patient exposure to ionizing radiation.
The application of deep learning for US image segmentation has been hampered due to the complexity of the modality and the lack of annotated training data, which is required for good DNN performance.
In order to facilitate the applications of US segmentation for automated AAA scanning, without the use of external imaging, an intermediate representation (IR) is required between US and CT so that CT labels and pretrained networks can be applied to the task of ultrasound image segmentation.

\subsubsection{Contributions}

We propose Common Anatomical CT-US Space (CACTUSS) which is an anatomical IR and is modality agnostic. The proposed method allows for:
1) real-time inference and segmentation of live ultrasound acquisitions, 2) training a deep neural network without the use of manually labeled ultrasound images,
3) reproducible and interpretable screening and monitoring of AAA.

We evaluate the results from the proposed approach by comparing it to a fully supervised segmentation network and investigate the use of the proposed method for measuring the anterior-posterior aortic diameter compared to the current clinical workflow. In total, the proposed method meets the clinical requirements associated with AAA screening and diagnosis.  The source code for our method is publicly available\footnote{\url{https://github.com/danivelikova/cactuss}}.

\section{Method and Experimental Setup} %~3 pages
CACTUSS (c.f. Fig.~\ref{fig:overview}) consists of three main phases: (1) Joint anatomical IR generator, (2) Domain Adaptation network, (3) Aorta Segmentation network.\\

\textbf{Ultrasound Simulation Parametrization: } 
Ultrasound Simulation has been an active research topic in the last two decades~\cite{treeby2011time,jensen2000fast,salehi2015patient}.
Since ultrasound data is limited in quantity and difficult to acquire, simulated ultrasound images are used to define an intermediate ultrasound representation.
To help define a common anatomical space, we take advantage of a hybrid US simulator introduced by~\cite{salehi2015patient}, implemented in ImFusion\footnote{ImFusion GmbH, Munich, Germany}.
In CACTUSS, the hybrid ray-tracing convolutional ultrasound simulator is used to define an anatomical IR with anisotropic properties, preserving the direction-dependent nature of US imaging while also having modality-specific artifacts of ultrasound imaging and well-defined contrast and resolution of CT.
This anatomical IR should reside on the joint domain boundary of US and CT distributions.
This has the benefit that from a single CT scan, a large number of samples of the IR can be created. 
Simulation parameters are listed in Table~\ref{tab:probe_params} and Table~\ref{tab:sim_params} and
describe the characteristics of the US machine, which allow for direct spatial mapping from the CT domain to the ultrasound domain.
Input to the simulator is a three-dimensional label map where each voxel is assigned six acoustic parameters that describe the tissue characteristics - speed of sound \textit{c}, acoustic impedance \textit{Z}, and attenuation coefficient $\alpha$, which are used to compute the acoustic intensity at each point along the travel path of the ray.
In this way, we create a virtual modality that provides important characteristics from ultrasound, such as tissue interfaces, while learning from annotated CT.
\begin{table}[h]
\centering
\makebox[0pt][c]{\parbox{0.95\textwidth}{%
    \begin{minipage}[b]{0.45\hsize}\centering
        \caption{Ultrasound scan parameters for ISS}
        \begin{tabular}{p{80pt}| p{50pt}}
            \thead{Parameter} & \thead{Value}\\ 
            \hline
            Probe width & 59mm \\
            Probe angle & $40^\circ$ \\
            Image depth & 100mm  \\
            Focus depth & 50mm  \\
            Scan lines & 196 \\
            Axial resolution & 1024  \\
            \hline
        \end{tabular}
        \label{tab:probe_params}
    \end{minipage}
    \hfill
    \begin{minipage}[b]{0.45\hsize}\centering
        \caption{Ultrasound simulation parameters for ISS}
        \begin{tabular}{p{80pt}| p{50pt}}
        \thead{Parameter} & \thead{Value}\\ 
        \hline
        Elevational Rays & 10  \\
        RF Noise & 0  \\
        Scale Exponent 1 & 1.0 \\
        Scale Exponent 2 & 0.2  \\
        TGC Alpha & 0.65  \\
        TGC Scale & 0.2  \\
        \hline        
        \end{tabular} 
        \label{tab:sim_params}
    \end{minipage}%
}}
\end{table}

\begin{figure}[t]
\includegraphics[width=\textwidth]{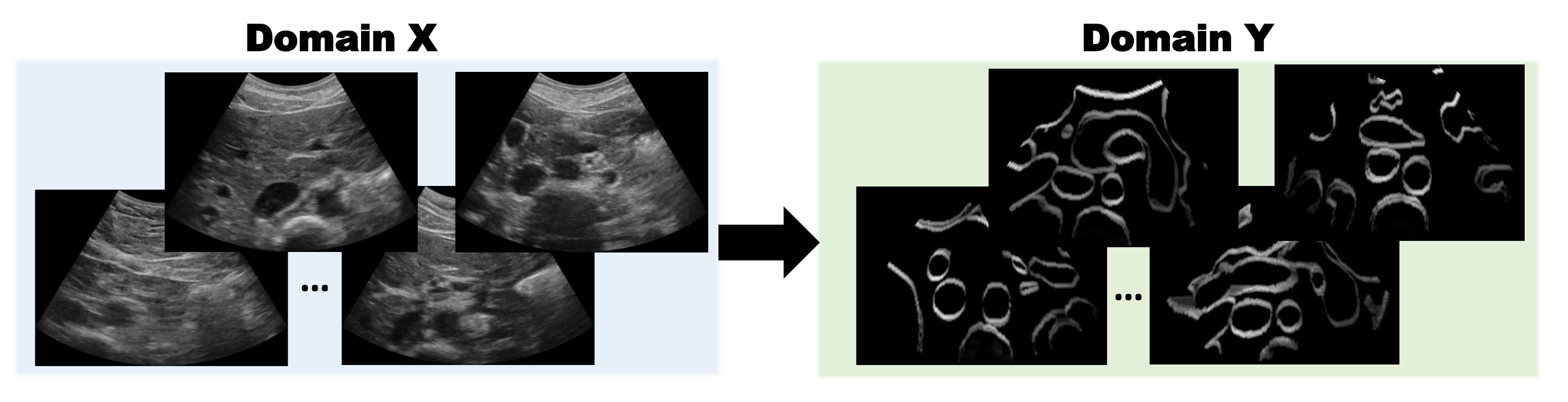}
\caption{Images from both image domains} \label{fig:domains}
\end{figure}

\textbf{Domain Adaptation }
Since there is a domain shift between the IR and real ultrasound B-modes, we learn a mapping between them while preserving the patient-specific anatomical characteristics of each image. 
In order to translate real ultrasound images into the IR 
we employ a recent Contrastive Learning for Unpaired image-to-image translation network (CUT)~\cite{park_contrastive_2020}.
The CUT network assumes a maximum correlation between the content information of a patch of the target image with the spatially corresponding patch of the source image vs. any other patches in the source image.
The network generator function $G : \mathcal{X} \mapsto \mathcal{Y}$ translates input domain images  $\mathcal{X}$ to look like an output domain images $\mathcal{Y}$, with unpaired samples from source $X = {x \in \mathcal{X}}$ and target $Y = {y \in \mathcal{Y}}$ respectively.
The generator $G$ is composed of an encoder $G_{enc}$ and a decoder $G_{dec}$, which are applied consecutively $\widehat{y} = G(z) = G_{dec}(G_{enc}(x))$. $G_{enc}$ is restricted to extracting content characteristics, while $G_{dec}$ learns to create the desired appearance using a patch contrastive loss~\cite{park_contrastive_2020}. The generated sample $Y$ is stylized, while preserving the structure of the input $x$.
Thus, samples can have the appearance of the IR while maintaining the anatomical content of the US image.

\textbf{Aorta Segmentation: } In the last phase, a segmentation network is trained on the samples from phase 1 to perform aorta segmentation on intermediate space images.
The corresponding labels can be directly extracted from the CT slices, saving manual labelling. Critically, no ultrasound image segmentation is required for CACTUSS.

\subsection{Data}
\label{sec:data}
Two domains of images are utilized in this work as can be seen in Figure~\ref{fig:domains}. \\
\textbf{Intermediate Space:} Eight partially labeled CT volumes of men and women were downloaded from a publicly available dataset Synapse\footnote{https://www.synapse.org/\#!Synapse:syn3193805/wiki/89480}.
These labels were augmented with labels of bones, fat, skin and lungs to complete the label map.
The CTs were used to generate 5000 simulated intermediate space samples with a size of 256x256 pixels. From those simulated data, a subset of 500 images was used for domain $Y$ for the CUT network training.
This dataset will be referred to as intermediate space set (ISS).\\
\textbf{In-vivo images:} Ten US abdominal sweeps were acquired of the aortas of ten volunteers (m:6/f:4), age = $26\pm3$ with a convex probe (CPCA19234r55) on a cQuest Cicada US scanner (Cephasonics, Santa Clara, CA, US).
Per sweep, 50 frames were randomly sampled, each with size 256x256 pixels, for a total of 500 samples and used for domain $X$ of the CUT network. 
For testing the segmentation network, which was trained only on IRs, %100 out of those 500 
a subset of 100 frames, with 10 random frames per volunteer
 was labelled by a medical expert and used as a test set. For the purpose of comparing against a supervised approach, additional images were annotated to train a patient-wise split 8-fold-cross validation network with 50 images per fold from 8 subjects.  
Additionally, 23 images from a volunteer not from the existing datasets were acquired from ACUSON Juniper (Siemens Healthineers, Erlangen, Germany) with a 5C1 convex probe and annotated for further evaluation.

\subsection{Training}
For phase 2 we train the CUT network for 70 epochs with a learning rate of $10^{-5}$ and default hyperparameters.
For phase 3 we train a U-Net~\cite{Unet} for 50 epochs with a learning rate of $10^{-3}$, batch size of 64, Adam optimizer and DSC loss.
Both models were implemented in PyTorch 1.8.1 and trained on a Nvidia GeForce RTX 3090 using Polyaxon\footnote{https://polyaxon.com/}.
Phase 3 training includes augmentations with rotation, translation, scaling and noise and is randomly split in 80-20\% ratio for training and validation, respectively.
For testing, the test set, consisting of 100 images from the in-vivo images, is inferred through the CUT network and translated into the common anatomical representation before being inferred with the phase 3 network.

\subsection{Evaluation Metrics} We use the following metrics to quantitatively evaluate our method:
For CUT we use the F\'rechet inception distance (FID)~\cite{FID} for performance evaluation and early stopping regularization.
FID quantifies the difference in feature distribution between two sets of images e.g. real and IR, using feature vectors from the Inception network.
As proposed in~\cite{FID} we use the second layer for FID calculation and consider the epochs with the top 3 FID scores and qualitatively select based on the desired appearance.
For the segmentation model, we report the average Dice Score (DSC) and mean absolute error (MAE) of the diameter of the resulting segmentation as proposed in~\cite{milletari2016v}.

\begin{figure}[t]
\centering
\includegraphics[width= \textwidth]{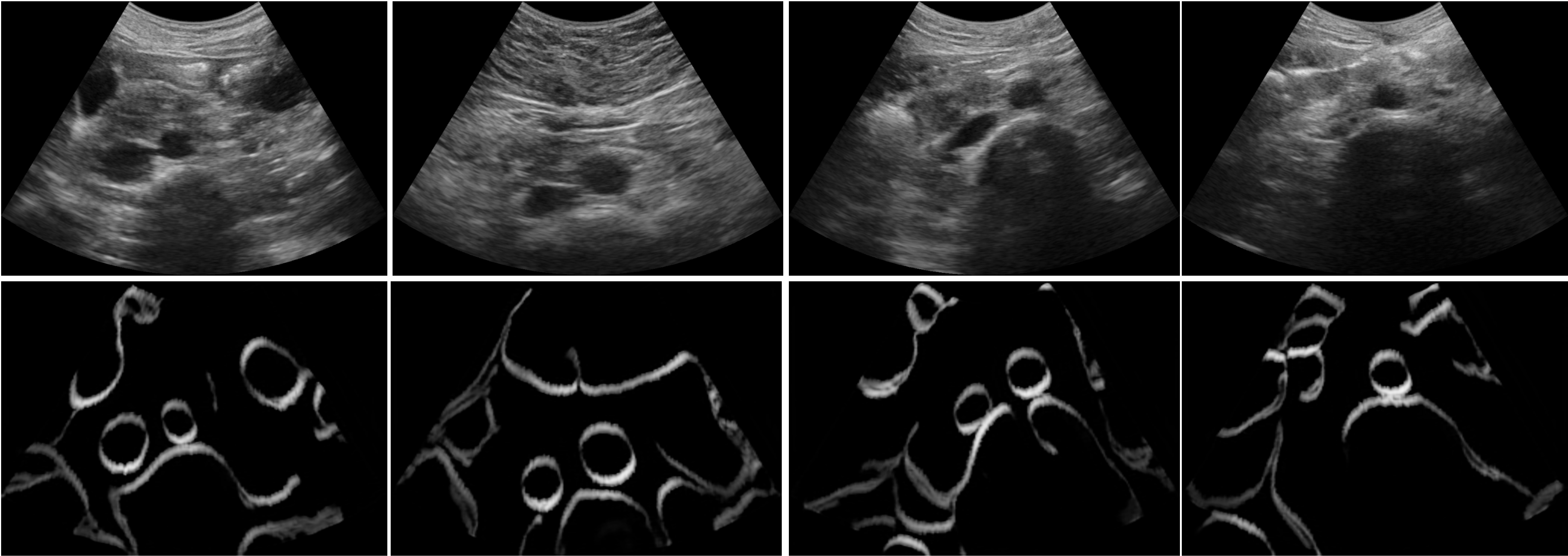}
\caption{Examples of B-mode images after inference through the CUT network to the IR. Top row: input B-mode. Bottom row: result after inference.} \label{fig:labelmap}
\end{figure}
\subsection{Experiments}
We test the proposed framework quantitatively wrt. the following experiments: \\ 
\textbf{Imaging Metrics:} We evaluate the accuracy of the proposed method by comparing it to a supervised network. For this, we train an 8-fold cross-validation U-Net, where each fold contains 50 in-vivo images from one subject. We test on 3 hold-out subjects and
report the average DSC. \\
\textbf{Clinical applicability:} 
We measure the anterior-posterior diameter of the aorta in mm, according to current clinical practice~\cite{HARTSHORNE2011195}, and report the
MAE and standard deviation compared to ground truth labels for both CACTUSS and the supervised segmentation. Clinically, an 
error of less than 8~mm is considered acceptable for a medical diagnosis of AAA~\cite{HARTSHORNE2011195}.
\\ \textbf{Robustness:} We evaluate against images of a patient scanned with a second  US machine as described in Section~\ref{sec:data}.
Thus we show how robust is the method to domain shift and again evaluate against the supervised network. 
\\ \textbf{Different Intermediate Representation:}  
We replace the proposed common anatomical IR with two alternatives to test the sensitivity of the proposed method's IR choice and specification. The first alternative tested processes CT slices with a Canny edge detector, bilaterial filter, and subsequent convex mask with shape from a convex US probe. The second is a realistic ultrasound simulation from the same label map as the ISS. These alternative IRs were evaluated on a DCS score on 100 in-vivo frames passed through the trained model, with expert annotation ground truth.

\section{Results and Discussion} 
In Tables~\ref{tab:comp_fully_sup} and~\ref{tab:secon_us_machine}, we present the DSC and MAE values of CACTUSS and a supervised U-Net when evaluated on the Cephasonics and Siemens scanners, respectively.
This evaluation is performed on two real-world scanners, while the CACTUSS phase 3 segmentation network has only been trained on synthetic samples from the ISS.
Remarkably, on the Cephasonics scanner, CACTUSS achieves a higher DSC in aortic segmentation and lower mean absolute error of aorta diameter measurements, a key metric in AAA diagnosis.
On the Siemens scanner, CACTUSS has a slightly higher MAE, but still exhibits a lower standard deviation. Furthermore, CACTUSS diameter measurement results are still within the clinically accepted range.
For this particular experiment, four out of 23 images were wrongly predicted and only from the supervised method.
\begin{table}
\centering
\makebox[0pt][c]{\parbox{0.99\textwidth}{%
    \begin{minipage}[b]{0.48\hsize} \centering
    \caption{Evaluation of CACTUSS on Cicada samples.} 
    \begin{tabular}{p{30pt}| p{60pt} | p{60pt} }
                    &  \thead{Supervised}          &\thead{CACTUSS} \\ 
        \hline
            DSC     & $85.7\pm0.02$   & $\mathbf{90.4\pm0.003}$\\
            MAE     & $4.3\pm1.9$     &  $\mathbf{2.9\pm1.9}$\\
     %\hline
    \end{tabular}
    % \vspace{5pt}
    \label{tab:comp_fully_sup}
    \end{minipage}
    \hfill
    \begin{minipage}[b]{0.48\hsize} \centering
    \caption{Evaluation of CACTUSS on Juniper samples.}
         \begin{tabular}{p{30pt}| p{60pt} | p{60pt} }
                    &  \thead{Supervised}          & \thead{CACTUSS}\\ 
            \hline
                DSC & 81.3 & \textbf{88.0}\\
                MAE &  $\mathbf{4.9\pm7.0}$ & $7.6\pm1.5$\\
     %\hline
    \end{tabular}
    % \vspace{5pt}
    \label{tab:secon_us_machine}
    \end{minipage}
    }}
    \end{table}

\textbf{Alternative Intermediate Representations:} 
Result from evaluating alternative IRs on the are reported in Table~\ref{tab:diff_interm_spaces}.
The proposed IR in CACTUSS outperforms both edge detection and realistic simulated ultrasound images.
\begin{table}[hb]
\centering
    \caption{Comparison of DSC of segmentation given alternative IRs.} 
    \begin{tabular}{ p{30pt}| p{70pt} | p{70pt} | p{70pt} }
      & \thead{Edges} & \thead{US Simulation} & \thead{CACTUSS}\\ 
     \hline
     DSC & 66.1 & 72.7 & \textbf{89.3}\\
     %\hline
    \end{tabular}
    \label{tab:diff_interm_spaces}
\end{table}

\subsection{Discussion}
CACTUSS was able to not only successfully segment real B-mode images while being trained only on IR data but was able to surpass the supervised U-Net as depicted in Table~\ref{tab:comp_fully_sup}. Furthermore, CACTUSS was able to achieve an aortic diameter measure accuracy of $2.9 \pm 1.9$  compared to $4.3 \pm 1.9$ for the supervised U-Net on the Cephasonics machine.
For both ultrasound devices, the diameter accuracy is within the accuracy required for clinical diagnosis AAA~\cite{HARTSHORNE2011195}. The results showed that it performed well independently from the machine used; however, the performance may vary due to different preprocessing steps and filters in each US machine.

Our testing of alternative intermediate representations showed the unique advantage that CACTUSS offers.
By embedding the anatomical layout in a space between US and CT, i.e. with the contrast of CT and attenuation and reflectively of ultrasound, the greatest segmentation performance was displayed.
By testing ultrasound simulation representations and an edge detection representation with high contrast, we show that the choice of representation is not arbitrary.
In the case of the US simulations, the reduction of performance is likely due to the increased complexity and the lower SNR of the image due to the addition of ultrasound-specific features such as shadows, reflections and speckle. Alternative representations are also possible, but our testing shows that including fundamental physical assumptions of both modalities enhances model performance.

One challenge of using CUT for domain adaptation is the possibility of hallucinations. Those networks are prone to hallucinate incorrect characteristics with the increasing number of training loops. However, we mitigate this issue by integrating FID to select the most performant CUT model. This approach can remain challenging for complex outputs; however, the CACTUSS IR is simplified in structure and is cleared from features such as speckle noise or reflections, thus improving trainability.

The reproducibility of diagnostic measurements between sonographers,
which is heavily dependent on their expertise, can lead to large inter as well as intraobserver variability. In particular, the differences in measurements between sonographers lie between 0.30-0.42cm, and the mean repeatability among technicians is 0.20cm~\cite{HARTSHORNE2011195}.
Neural network-based methods provide a standardized computer-aided diagnostic approach that improves the reproducibility of results since models are deterministic. 
In this way, CACTUSS shows reproducible deterministic results, which are within the clinically accepted ranges, and shows stability in evaluation results.

Additionally, CACTUSS shows reproducible results on images from different US machines, which is a positive indication that the algorithm can be machine agnostic. Moreover, CACTUSS can also be referred to as modality agnostic since intermediate representation images can also be generated from other medical modalities such as MRI.
Initial experimental results on AAA sample images show that CACTUSS is able to successfully generate an IR for AAA B-mode images independently of anatomical size and shape\footnote{\url{https://github.com/danivelikova/cactuss}}. 
The desired segmentation performance can be achieved by re-training the segmentation network on any in-distribution data. This shows that CACTUSS is applicable to AAA cases and has the potential to generalize to other applications and anatomies. This demonstrates the adaptivity of the proposed method.

\section{Conclusion} 
In this work, we presented CACTUSS, a common anatomical CT-US space, which is generated out of physics-based simulation systems to address better the task of aorta segmentation for AAA screening and monitoring.
We successfully show that US segmentation networks can be trained with existing labeled data from other modalities and effectively solve clinical problems.
By better utilizing medical data, we show that the problem of aorta segmentation for AAA screening can be performed within the rigorous standards of current medical practice.
Furthermore, we show the robustness of this work by evaluating CACTUSS on data from multiple scanners.
Future work includes the integration of CACTUSS in robotic ultrasound platforms for automatic AAA screening and clinical feasibility studies of the method.

\subsubsection{Acknowledgements} 

This work was partially supported by the department of Interventional Radiology in Klinikum rechts der Isar, Munich. We would also like to thank Teresa Sch\"afer for helping us with the data annotations and Dr. Magdalini Paschali for revising the manuscript.

\bibliographystyle{splncs04}
\bibliography{refs.bib}

\end{document}